\documentstyle[twocolumn,aps]{revtex}
\input psfig

\begin{document}
\draft

\twocolumn[\hsize\textwidth\columnwidth\hsize\csname @twocolumnfalse\endcsname

\title{Spectral Functions of One-dimensional Models of Correlated Electrons}
\author{
  Julien Favand$^{a}$, Stephan Haas$^{b}$, Karlo Penc$^{c}$\cite{*}, 
  Fr\'ed\'eric Mila$^{a}$, Elbio Dagotto$^{d}$ 
  }

\address{
     $(a)$ Laboratoire de Physique Quantique, Universit\'e Paul Sabatier,
     31062 Toulouse (France) \\
     $(b)$ Theoretische Physik, ETH-H\"onggerberg, 8093 Z\"urich (Switzerland) \\
     $(c)$ Max-Planck-Institut f\"ur Physik komplexer Systeme, 
          Bayreuther Str.~40, 01187 Dresden (Germany) \\
     $(d)$ Department of Physics and National High Magnetic Field Laboratory,
     Florida State University, Tallahassee, FL 32306 (USA).
}

\maketitle

\begin{abstract}
Using the Ogata-Shiba wave function, the spectral functions of the
one-dimensional (1D) infinite $U$ Hubbard model are calculated 
for various concentrations. 
It is shown that the ``shadow band'' feature due to $2k_F$
fluctuations becomes more intense  close to half-filling.
Comparing these results 
with exact diagonalization 
data obtained on finite clusters for the finite $U$ Hubbard model and for 
the $t-J$ model, it is also shown that this feature remains well-defined 
for physically reasonable values of the parameters ($U/t\simeq 10$, $J/t\simeq
0.4$). The ``shadow'' structure in the spectral functions 
 should thus be observable in angle-resolved photoemission experiments 
for a variety of quasi-one dimensional compounds.
\end{abstract}


\vskip2pc]

\narrowtext
Spectral functions are very useful to understand the electronic structure of
solids. They are defined by 
\begin{eqnarray}
  A(k,\omega) &=&
 \sum_{f,\sigma} 
  \left| \langle f,N \!+\! 1| c^\dagger_{k,\sigma} |0,N\rangle \right|^2
  \delta(\omega \!-\! E^{N+1}_f \!+\! E^N_0) 
  \>,\nonumber\\
  B(k,\omega) &=&
 \sum_{f,\sigma} 
  \left| \langle f,N \!-\! 1| c^{\phantom{\dagger}}_{k,\sigma} |0,N\rangle \right|^2
    \delta(\omega \!-\! E^{N}_0 \!+\! E^{N-1}_f) 
  \>,
\nonumber
\end{eqnarray}
in the  standard notation,
and in principle they can 
be measured in angular-resolved inverse photoemission and 
photoemission experiments, respectively. 
The effect of electron-electron interactions on the spectral functions of
the three-dimensional Coulomb gas has been
investigated in much detail several years
ago\cite{hedin}. With respect to the simple $\delta(\omega - \varepsilon_k)$ 
structure of the
non-interacting case, the main differences are: i) A shift of the energy of the 
quasi-particle band; ii) A broadening of the quasi-particle peak; iii) A 
renormalization of the weight of the
quasi-particle peak, compensated by the appearance of an incoherent background;
iv) The presence of another band - a plasmon band - due to 
the long-range nature of
the Coulomb potential. For a short-range repulsion, such as 
the on-site interaction
of the Hubbard model, there is no plasmon band, and the spectral function is
expected to have only one well-defined feature, the quasi-particle band, on top
of an incoherent background. The validity  of this simple 
picture in lower dimensional systems 
is currently under intense discussion. In two dimensions(2D), 
there are at least two important issues:
1. the actual existence of a quasiparticle peak, and 2. the generation by
antiferromagnetic short-range fluctuations close to half-filling
of additional well-defined features 
in the spectral functions, 
called generically ``shadow bands''\cite{kampf}.
The absence 
of exact results in 2D for correlated models makes the
interpretation of the experimental results quite difficult. For high--T$_c$
cuprates, these issues are  still controversial\cite{2D}.

The situation is much more favorable in 1D. First of all, it is known 
that there is no quasi-particle peak 
in the spectral function but two divergences 
due to spin-charge
separation\cite{meden}. Besides, using the Ogata-Shiba
wave-function\cite{shiba}, exact results have been  
recently obtained\cite{penc}
for the spectral
functions of the infinite $U$ limit of the Hubbard model defined by:
\begin{equation}
  {\cal H}= -t \sum_{i,\sigma} 
   \left(
    c^\dagger_{i,\sigma} c^{\phantom{\dagger}}_{i+1,\sigma}
   + h.c. \right)
   + U \sum_{i} n_{i,\uparrow} n_{i,\downarrow},
\end{equation}
or, equivalently, for the $J=0$ limit of the $t-J$ model defined by:
\begin{eqnarray}
  {\cal H} & = & -t \sum_{i,\sigma} 
   \left(
    \tilde c^\dagger_{i,\sigma} \tilde c^{\phantom{\dagger}}_{i+1,\sigma}
   + h.c. \right)\nonumber \\ 
& + & J  \sum_i 
    (\vec{S_i}. \vec{S}_{i+1}-{1\over 4}n_i n_{i+1}),
\end{eqnarray}
where $\tilde c$ are the 
usual hole operators, and the rest of the
notation is standard. Combining these results with
numerical simulations should allow us to reach a precise picture of the
spectral functions of one-dimensional systems.

In this paper we will concentrate our efforts on three aspects of
the spectral functions of 1D systems: 1. 
The first one is the extra weight created
near the Fermi energy close to half-filling by short-range 
antiferromagnetic correlations. It is this feature that is 
usually referred to as ``shadow band'' in the literature. 2. The 
second issue presented here
is that the ``shadow'' weight is actually
part of an extended band that can be followed in the 
whole
Brillouin zone. 3. The third point 
is that this extended band still exists away from
half-filling due to diverging $2k_F$ fluctuations, although its weight near the
Fermi level diminishes rapidly with hole doping.
In portions of the paper we will refer to all these features 
with the common language of ``shadow band'',
but it must be clear to the reader that in doing so we are extending the
usual meaning of this term to include all the
extra interesting features caused by $2k_F$ fluctuations, both close and
away from the Fermi energy, that
appear in the spectral functions of 1D systems. 

The presence of shadow bands in 1D systems 
was first revealed in studies of the 
spectral function corresponding to a hole injected into one
dimensional Heisenberg models at half-filling\cite{haas}. They were
identified as potentially observable features in photoemission
experiments for CuGeO$_3$ and other undoped compounds where antiferromagnetic
fluctuations are important. Other related aspects 
were first discussed in Ref.\cite{penc} where
a calculation of the spectral functions in the infinite $U$ limit showed
unambiguously that, in addition to the charge and spin features typical of a
Luttinger liquid, there is a well-defined 
structure whose dispersion can be followed in the whole Brillouin zone.
This accumulation of weight in the spectral functions
originates from diverging spin fluctuations\cite{com} at $2k_F$, the 1D
equivalent
of the diverging fluctuations at ($\pi/a$,$\pi/a$) in the 2D case at
half-filling, 
and by analogy it was also called a shadow band.

A clear identification of these features in photoemission experimental 
results for  quasi-one dimensional systems is still lacking. 
On the theoretical side, progress can be made along two lines: First,
by estimating the evolution of the shadow band features, 
in particular their intensity, 
with electron concentration to determine which systems are a priori the best 
candidates to 
observe the effect. Second, by extending the
calculations to more realistic parameters, i.e. $U< + \infty$ or $J>0$, to
investigate if
the band remains a well-defined structure when the 
on-site repulsion is no longer infinite. 

To address the first point, we have calculated the spectral functions for an
infinite repulsion along the lines of Ref.\cite{penc} for various
concentrations\cite{sorella}. Some results for $B(k,\omega)$ are shown in 
Figure 1. The evolution between
$n=1/2$ and $n=1$ is smooth. When $3k_F$ is larger than $\pi/a$, i.e. for $n>2/3$,
the band at finite binding energy discussed in Ref.\cite{penc} at $n=1/2$
is now folded back when it hits the zone boundary and it crosses 
the Fermi level at $2\pi/a-3k_F$. This detail is very important since now
non-trivial structure in the spectral functions appears near the Fermi energy.
The rest of the shadow band evolves smoothly with density becoming
more intense close to half-filling. Note, however, that the effect is
likely less dramatic as expected in 2D systems for realistic values of $U/t$. 
This is due to the fact that in 1D there
is always perfect nesting, independent of the band-filling, since the 
Fermi
surface consists of only two points. The 1D best candidates to show these
shadow band effects are compounds with a
concentration close to 1, as anticipated in Ref. \cite{haas}. However, 
quarter-filled systems should not be disregarded since
the portion of the band that appears at energies between
$-2t$ and $-t$ for $0\le k \le \pi/2a$ could be experimentally
detectable, unless the usual
large spurious backgrounds that contaminate photoemission 
results as we move away from the Fermi level hide the
signal in the noise. 

\begin{figure}[hp]
\centerline{\psfig{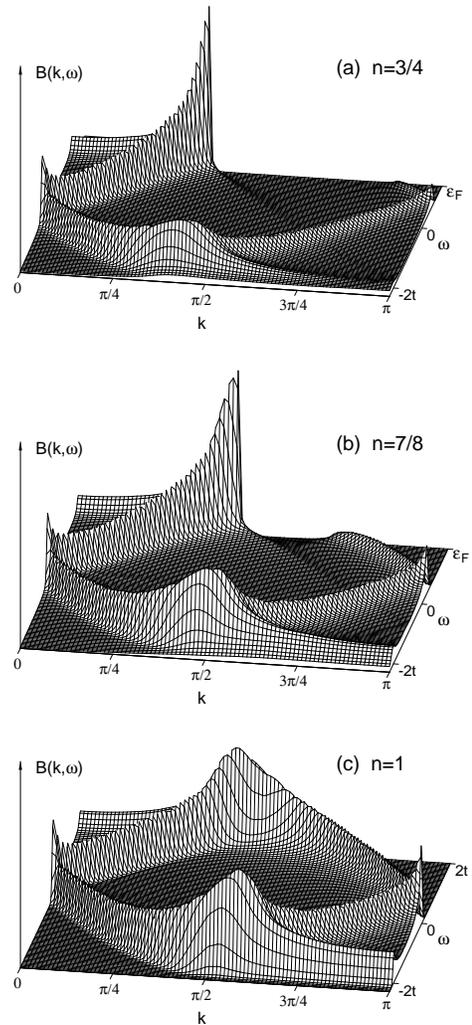}}
\vspace{0.5cm}
\caption{
Spectral function $B(k,\omega)$ of the $U\rightarrow +\infty$ Hubbard model
calculated with
the help of the Ogata-Shiba wave-function. a) $n=3/4$ ($k_F=3\pi/8$), 
200 sites; 
b) $n=7/8$ ($k_F=7\pi/16$),
176 sites; c) $n=1$ ($k_F=\pi/2$), 190 sites.}
\label{fig:3d}
\end{figure}

The second point, i.e. calculations for realistic parameters, 
is more difficult to address. Even for models that are
integrable, such as the Hubbard model or the 
super-symmetric $t-J$ model ($J/t=2$),
a calculation of the spectral functions for large systems remains an open 
problem. The only available methods are numerical\cite{dagotto}. 
Relatively large systems can
be handled with Monte Carlo simulations\cite{montecarlo}, 
but the results require an analytic
continuation which is difficult to control. Thus exact diagonalizations of
finite clusters seem to be more appropriate. 
Such calculations have already been
presented for 1D systems\cite{haas,maekawa,kim}. The main limitation of these
simulations is the size of the systems - typically 16 to 20 sites close to
half-filling. The spectral functions obtained with the exact diagonalization
method consist of a collection of discrete $\delta$ 
functions, and they are difficult to interpret without an underlying theory.

The originality of the present work is to use the essentially exact results 
obtained for 
infinite $U$ with the Ogata-Shiba wave function to interpret the results for 
finite clusters.
To illustrate how this works, we have compared in Fig. 2 the ``exact'' results 
for $n=7/8$ (Fig. 2a) with 
the results obtained for the same model on a 16 site cluster
using exact diagonalizations (Fig. 2b). Details about the numerical method can
be found in Ref.\cite{dagotto}. Clearly, the
calculation on 16 sites is not able to reproduce a continuum:
The true spectral function has a continuum between the peaks corresponding to
the charge and spin excitations of the Luttinger liquid\cite{meden},
whereas the spectral function of Fig. 2b
has several peaks in the  

\begin{figure}[hp]
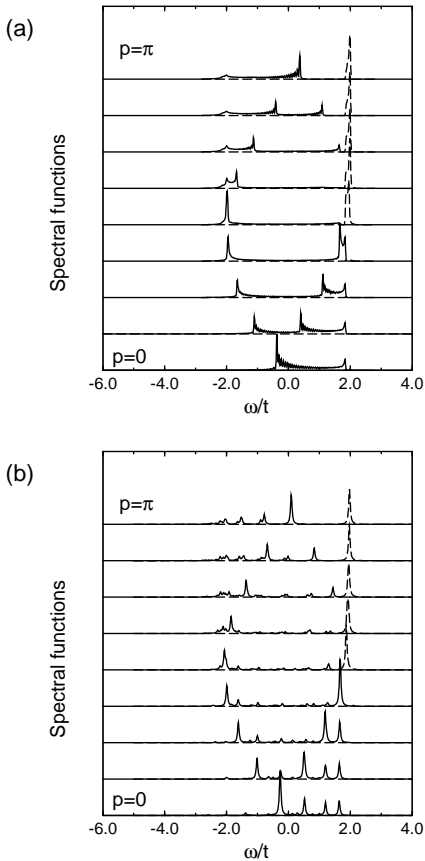

\centerline{\psfig{figure=fred2.epsi,width=4.9cm,angle=270}}
\vspace{0.5cm}
\centerline{\psfig{figure=fred3.epsi,width=4.9cm,angle=270}}
\vspace{0.5cm}
\caption{
Spectral functions of the $U\rightarrow +\infty$ Hubbard model at
$n=7/8$
for some selected momenta. Solid lines: $B(k,\omega)$; Dashed lines:
$A(k,\omega)$. a) "Exact" results obtained with the Ogata-Shiba
wave-function for 176 sites. The rapid oscillations are finite-size effects;
b) Numerical results obtained on a 16 site cluster using exact
diagonalizations and periodic boundary conditions.
The $\delta$ peaks have been given an arbitrary broadening of
$0.03 t$.}
\label{fig:2d}
\end{figure}

\noindent 
same energy interval. This is best observed
for $k=0$, where 
the continuum is 
replaced by a collection of 4 peaks. This proves
that detecting the position of the spin and charge
excitations corresponding to
the Luttinger liquid features is not possible on the basis of exact
diagonalizations of small clusters, unless a variety of boundary conditions is
used and a finite size scaling analysis is carried out. In spite of 
this problem, the shadow band is clearly
visible on the data of Fig.2b, and both its intensity and its 
dispersion are in
good agreement with the exact results. The continuum that exists 
between the Van
Hove singularity at $-2t$ and the shadow band around $k=\pi$ is again poorly 
described on the
finite cluster, but the main band and the shadow band
can be clearly identified. Thus we believe that discussing the intensity of the
shadow band on the basis of exact diagonalization results is meaningful.

\begin{figure}[hp]
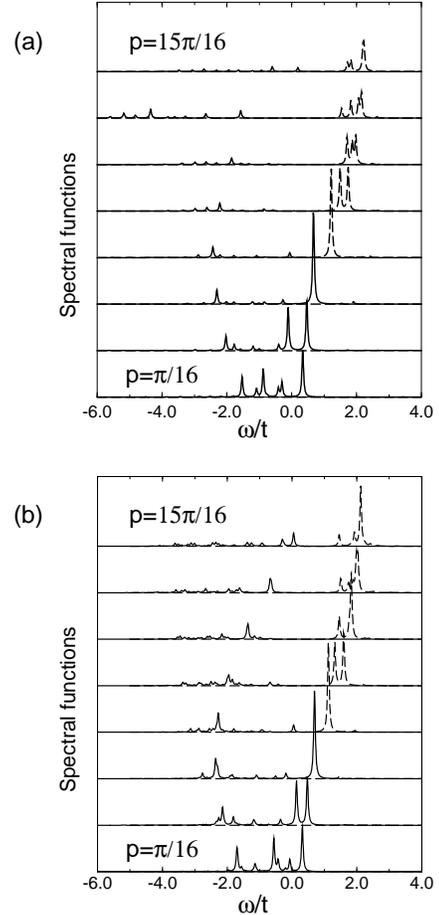

\centerline{\psfig{figure=fred4.epsi,width=4.9cm,angle=270}}
\vspace{0.5cm}
\centerline{\psfig{figure=fred5.epsi,width=4.9cm,angle=270}}
\vspace{0.5cm}
\caption{
Spectral functions for the Hubbard and $t-J$ model at $n=3/4$
obtained on a 16
site cluster using exact
diagonalizations and antiperiodic boundary conditions. Solid lines:
$B(k,\omega)$; Dashed lines:
$A(k,\omega)$.
a) Hubbard model with $U/t=10$; b) $t-J$ model with $J/t=0.4$.}
\label{fig:XYcut}
\end{figure}

Let us now turn to more realistic parameters. To make our analysis below
more convincing, we have calculated the spectral functions 
for both the Hubbard model
and the $t-J$ model, and we have compared the results using the relation
$J=4t^2/U$. 
Some results obtained at $n=3/4$ for $U/t=10$ and $J/t=0.4$ are depicted on 
Fig. 3a and 3b, respectively. The results are roughly the same.
 In agreement with what was stated before, we will not try to interpret the
details of the fine
structure around $\omega/t=0$ for small momentum or around $\omega/t=-2$ for 
large momentum. What is clearly
observable, although it carries a small weight, are ``shadow'' features 
located at the same position
as in the infinite $U$ case. Its
intensity is larger in the $t-J$ case than in the Hubbard case, 
and this difference grows
as $U$ becomes smaller (since the agreement between Hubbard
and $t-J$ models is expected only at large $U/t$). 
Increasing $J/t$ does not
suppress the spectral weight around the shadow band but simply makes the
feature broader, while decreasing $U/t$ suppresses the intensity of the shadow
band quite dramatically close to the zone boundary. The spectral weight which
is lost appears around $\omega = U$, a region not 
shown on Fig. 3b for clarity. 
This likely arises from corrections of order $1/U$ that are not included in
the $t-J$ model\cite{eskes}.
However, we have checked that, for parameters as small as $U/t=5$, the shadow 
band remains quite
intense for small wave vectors.

\begin{figure}[hp]
\centerline{\psfig{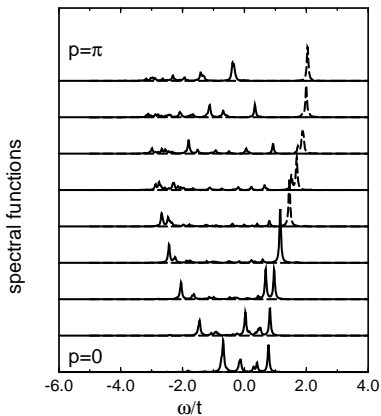}}
\vspace{0.5cm}
\caption{
Spectral functions for the $t-J$ model at $n=7/8$
obtained on a 16
site cluster using exact
diagonalizations and periodic boundary conditions. Solid lines:
$B(k,\omega)$; Dashed lines:
$A(k,\omega)$.}
\end{figure}

Upon approaching half-filling, the shadow band becomes quite intense. 
For instance, we show in Fig. 4 the spectral
functions for $J/t=0.4$ at $n=7/8$. A photoemission
experiment performed on a compound described by such parameters should be able
to detect the portion of the
shadow band the closest to the Fermi energy, 
and actually follow it in a large part of the Brillouin zone.

In conclusion, we have presented a detailed study of the ``shadow'' band that 
appears in the spectral
functions of 1D systems due to diverging $2k_F$
fluctuations\cite{penc}. 
First, we have studied its evolution with electron concentration.
As anticipated in Ref. \cite{haas}, its weight near the Fermi level is
considerably enhanced close to half-filling. Then, we have shown that this
feature is 
quite robust for realistic values of $U/t$ (or increasing $J/t$ ). 
This is true even away from
half-filling specially regarding the weight located at a finite binding
energy\cite{penc}. 
We believe that it should be possible to detect
this band  experimentally either close
to the Fermi level in nearly half-filled systems or at energies of the order of
the band width for systems far from half-filling, and
it is our hope that the present study 
will encourage experimentalists to search for this novel
feature of the spectral functions of 1D correlated electron systems.

We thank D. Poilblanc and H. Shiba
for useful discussions. We also thank IDRIS (Orsay) for allocation of CPU time
on the C94 and C98 CRAY supercomputers. E. D. is supported by grant
NSF-DMR-9520776.


\begin{references}
\bibitem{com} In principle, it can also originate from the density 
fluctuations at $2k_F$. However, in the $U\rightarrow +\infty$ limit, 
the density
$2k_F$ fluctuations are absent: They appear at finite $U$ only.
\end{references}
\end{document}